\documentclass[9pt,journal,twoside,web]{ieeecolor}

\usepackage{subcaption}
\usepackage{generic}

\usepackage[T1]{fontenc}
\usepackage[utf8]{inputenc}
\usepackage[english]{babel}

\makeatletter
\adddialect\l@ENGLISH\l@english
\makeatother

\usepackage{array}
\usepackage{units}

\usepackage{amsmath,amssymb,amsfonts}
\usepackage{esint}

\usepackage{xcolor}
\usepackage{cite}
\usepackage{graphicx}
\usepackage{epstopdf}

\usepackage{algorithm}
\usepackage{algorithmic}

\usepackage{textcomp}

\usepackage{hyperref}
\hypersetup{hidelinks}

\allowdisplaybreaks

\newtheorem{remark}{Remark}
\newtheorem{theorem}{Theorem}
\newtheorem{assu}{Assumption}

\usepackage{etoolbox}

\patchcmd{\thebibliography}
  {\section*{\color{black}}}
  {\section*{\color{black}References}}
  {}
  {\PackageWarning{bibheading}{Bibliography heading patch failed}}

\begin{document}
\title{Backstepping-Forwarding Design for Hyperbolic PIDE-ODE Systems with Distributed Interconnection and an Arbitrarily Large Input Delay}
\author{Yidan Cao, Xiang Xu, \IEEEmembership{Senior Member, IEEE}, Lu Liu, \IEEEmembership{Senior Member, IEEE}
\thanks{This work was supported by the National Natural Science Foundation of China under Grants 62373314 and the Research Grants Council of the Hong Kong Special Administrative Region of China under Grant CityU/11210424. The material in this paper was not presented at any conference. (Corresponding author: Lu Liu) }
\thanks{Yidan Cao and Lu Liu are with the Department of Mechanical Engineering, City University of Hong Kong, Kowloon, Hong Kong, SAR, China (e-mail: yidancao4-c@my.cityu.edu.hk, luliu45@cityu.edu.hk).}
\thanks{Xiang Xu is with School of Automation and Intelligent Sensing, Shanghai Jiao Tong University, Shanghai 200030, China (e-mail: xiangxu5@sjtu.edu.cn). }
}
\maketitle

\begin{abstract}
This paper studies first-order hyperbolic partial integro-differential equation (PIDE) and ordinary differential equation (ODE) cascades with distributed interconnection and an arbitrarily large known constant input delay. Within the backstepping--forwarding framework, a forwarding-type transformation is designed to reduce the distributed interconnection to an equivalent boundary-coupled Dirichlet cascade  without requiring the PIDE to be open-loop stable. A subsequent backstepping design stabilizes the PIDE and the ODE while compensating for the input delay. Incompatible boundary conditions in the delay-compensation kernel equations may produce discontinuities along characteristic lines.  We determine their jump locations and amplitudes and prove that the interface terms cancel when deriving the target equations, so the kernel discontinuities do not change the target system.
Closed-loop exponential stability is established without additional plant restrictions to ensure kernel continuity.
A numerical example illustrates the effectiveness of the proposed approach.
\end{abstract}

\begin{IEEEkeywords}
Backstepping-forwarding framework, distributed interconnection, discontinuous kernel functions, arbitrarily large input delay, hyperbolic PIDE-ODE system, boundary control
\end{IEEEkeywords}

\section{Introduction}
\label{sec:introduction}
\IEEEPARstart{P}{artial} differential equations (PDEs) are widely used to describe spatially distributed processes in engineering. In many applications, these processes interact with finite-dimensional components governed by ordinary differential equations (ODEs), giving rise to coupled PDE--ODE systems. 
Hyperbolic PDE-ODE systems have wide applications, such as drilling stick-slip \cite{sagert2013backstepping}, systems of open channels and transmission lines \cite{bastin2016stability}, vehicular traffic flow \cite{goatin2006aw}, pipeline gas transport \cite{gugat2011gas}, communication networks \cite{espitia2017dynamic}, and deep-sea drilling systems \cite{saldivar2014overview, wang2020adaptive}. Building upon this broad foundation of hyperbolic PDE-ODE application background, significant research in recent years has focused on the boundary control of hyperbolic PDE-ODE systems.

PDE--ODE interconnections may involve boundary values or distributed integrals of the PDE state.
While existing research on hyperbolic PDE-ODE systems has predominantly addressed Dirichlet coupled configurations \cite{hasan2016boundary,di2018stabilization,deutscher2018output,deutscher2019output,redaud2024output,auriol2018delay,krstic2008backstepping}, distributed interconnections characterized by interactions throughout the spatial domain still require further exploration.  Such distributed dynamics arise when the PDE serves as a spatially distributed input to the ODE, or when the ODE state acts as a distributed source term within the PDE. 
The backstepping method, although effective for Dirichlet coupling and for certain delay-compensated systems  \cite{krstic2008backstepping}, is  limited to strict-feedback structures and is not directly applicable to distributed interconnections.

Forwarding transformations can complement backstepping by modifying the ODE state to account for distributed PDE coupling. 
For diffusion and counter-convection actuator dynamics, this approach has been used to compensate  for the distributed influence on an ODE~\cite{bekiaris2010compensating}.
Other backstepping-based designs address robust output regulation for parabolic PIDEs with in-domain coupling~\cite{deutscher2017backstepping} and stabilization of parabolic PDE--ODE  subsystems~\cite{xu2023stabilization}.
For hyperbolic systems, backstepping--forwarding and Fredholm transformations have been used to handle internal nonlocal  couplings~\cite{bribiesca2015backstepping,redaud2022stabilizing}, while stabilization and regulation have also been studied for boundary-coupled PDE--ODE systems~\cite{di2018stabilization,redaud2024output}.
Distributed interconnections with delays were considered in~\cite{xu2024backstepping}, where a forwarding transformation removes the distributed PDE term from the ODE and predictor feedback compensates for the remaining distributed input delay. 
However, the stability analysis assumes that the homogeneous PDE is open-loop stable.
As a result, these results do not directly address the present setting of a distributed hyperbolic PIDE--ODE cascade with potentially unstable PIDE dynamics and a boundary input delay.
This motivates the construction of a forwarding-type integral transformation that reduces the distributed interconnection to an equivalent boundary-coupled cascade.
The reduction  leaves the PIDE dynamics unchanged and separates the treatment of the distributed interconnection from the subsequent stabilization and delay compensation steps. In this way, established boundary-control techniques can be applied after the distributed coupling has been converted into boundary coupling.

Delay compensation has been studied using backstepping for actuator and sensor delays~\cite{krstic2008backstepping} and using predictor feedback for distributed input delays~\cite{bekiaris2010lyapunov}. A broader review of backstepping methods for interconnected PDE systems is  given in~\cite{vazquez2026backstepping}. For parabolic systems, backstepping has been used to compensate long input delays in unstable reaction--diffusion PDEs~\cite{krstic2009control}, while spectral reduction and predictor feedback have been used to stabilize delayed-input ODE--heat cascades~\cite{lhachemi2025delayed}. For hyperbolic systems, existing results include delay-robust stabilization of boundary-coupled PDE--ODE systems~\cite{auriol2018delay}, compensation of uncertain input delays in coupled PDEs~\cite{zhang2023robust}, and delay-adaptive control of first-order PIDEs and PDE--ODE cascades~\cite{wang2024delay,wang2024delay2}. However, several of these designs impose additional restrictions on the original plant coefficients in order to avoid kernel boundary incompatibilities, including $g(1)=0$ in~\cite{wang2024delay} and $c(1)=0$ in~\cite{qi2024neural,qi2025neural}, where $g$ and $c$ denote the corresponding coefficients in those plant models. Therefore, the existing literature does not provide a systematic treatment of the kernel boundary incompatibilities that arise in the delayed hyperbolic PIDE--ODE cascade considered here. This motivates the present work, in which we analyze discontinuous kernels.

\begin{figure*}
    \centering
    \includegraphics[width=4in]{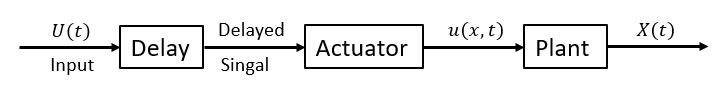}
    \caption{Hyperbolic PIDE-ODE system with an arbitrarily large input delay}
    \label{xitongjiegou}
\end{figure*}

Characteristic kernel jumps and their contributions to integral transformations have been analyzed for quasilinear hyperbolic systems~\cite{hu2019boundary}. Bounded kernels have been established for coupled linear hyperbolic systems~\cite{hu2015control}, and piecewise-continuous kernels have been used in Fredholm transformations for interconnected hyperbolic systems~\cite{redaud2022stabilizing}, while piecewise-differentiable kernels have been considered in backstepping designs for coupled parabolic systems~\cite{vazquez2016boundary}.
These works provide tools for handling discontinuous kernels, but they do not address the specific discontinuity mechanism that appears here. In the present setting, input-delay compensation generates two kernel boundary conditions that may be incompatible at a common corner. The resulting kernel must therefore be allowed to be discontinuous. We explicitly determine the jump locations and amplitudes and prove that the associated interface terms cancel in the target-system derivation, thereby preserving the transformed dynamics and closed-loop stability.

This paper focuses on first-order hyperbolic PIDE-ODE systems with distributed interconnections and an arbitrarily large input delay. 
The main contributions can be summarized as follows.

First, we construct a forwarding-type transformation that reduces the distributed interconnection to an equivalent boundary-coupled Dirichlet cascade without requiring open-loop stability of the actuator dynamics. Distributed interconnections of parabolic PDE-ODE systems have been studied in~\cite{bekiaris2010compensating} where the  PDEs are open-loop stable under the specified boundary conditions, while
\cite{xu2024backstepping} explicitly assumes stability of the homogeneous PDE. The hyperbolic design in~\cite{bribiesca2015backstepping} transforms the PIDE state to remove internal Fredholm integral terms, whereas our forwarding transformation reduces the distributed PIDE--ODE interconnection to boundary coupling.

Second, we construct discontinuous kernels to compensate for the input delay at the boundary of the hyperbolic PIDE, without imposing additional restrictions on the plant coefficients to ensure kernel continuity. Existing designs impose conditions such as $g(1)=0$ in~\cite{wang2024delay} and $c(1)=0$ in~\cite{qi2024neural,qi2025neural} to ensure compatibility of the kernel boundary conditions. We allow $m(1)\neq0$ and show that incompatible kernel boundary data generate jumps along characteristic lines of the kernel equations. We solve the kernel equations on each side of these characteristic interfaces and derive the jump amplitudes. 
The main technical difficulty is that the kernel jumps produce additional terms when the integral transformations are differentiated. We prove that interface terms cancel when deriving the target equations, so the discontinuous kernels do not introduce additional terms into the target system.
The kernel discontinuities arise from the propagation of incompatible boundary data along characteristics of the hyperbolic delay-compensation equations. This mechanism is inherent to the hyperbolic structure and does not depend on whether the PDE is cascaded with an ODE.

The rest of this paper is organized as follows. Section~\ref{2} formulates the problem and represents the input delay by a transport PDE. Section~\ref{3} presents the distributed-to-boundary reduction and backstepping design, analyzes the kernel discontinuities, and proves closed-loop exponential stability. Section~\ref{4} provides numerical simulations. Section~\ref{5} concludes the paper and discusses future work.

\newtheorem{notation}{Notation}
\begin{notation}
$\|\cdot\|$ denotes the Euclidean vector norm and the corresponding induced matrix norm, while $|\cdot|$ denotes the absolute value of a scalar. For a vector $\xi$ and a scalar function $\varphi$, we write
\begin{align}
\|\xi\|=\sqrt{\xi^\top\xi},
\qquad
\|\varphi\|_{L^2}
=\left(\int_0^1|\varphi(s)|^2\,ds\right)^{1/2}.
\end{align}
\end{notation}

\section{Problem Formulation}\label{2}
Motivated by the  PIDE models in~\cite{krstic2008backstepping,wang2024delay}, we consider a PIDE--ODE cascade system to study the combined effects of transport dynamics, distributed interconnection, and input delay on the stability of the entire coupled system, as shown in Fig.\ref{xitongjiegou}:
\begin{subequations}
\begin{align}
    &\dot{X}(t) =AX(t)+\int_{0}^{1} \mathcal{B}(x)u(x,t)dx,\label{xt} \\
    &u_{t}(x,t) = \lambda u_{x}(x,t)+m(x) u(0,t)+\int_{0}^{x} f(x,y)u(y,t)dy,\label{ut} \\
    &u(1,t) =U(t-D),\label{t-d}
\end{align}
\end{subequations}
where $X(t)\in\mathbb R^n$ is the ODE state, $u:[0,1]\times[0,+\infty)\to\mathbb R$ is the actuator PIDE state, and $U(t)\in\mathbb R$ is the control input, which is delayed by $D>0$ units of time. Throughout the paper, we assume that $\lambda>0$, $\mathcal B(x)$ is continuous on $[0,1]$, and $m\in C^{1}([0,1]), f\in C^{1}(\mathcal T)$, where $\mathcal T=\{(x,y)\in[0,1]^2:\ 0\le y\le x\le 1\}$. The analysis considers a scalar linear hyperbolic PIDE with constant transport speed and a Volterra integral term, coupled to a finite-dimensional ODE through a distributed interconnection. The PIDE dynamics are homodirectional, and the coupled system is controlled through a single effective input channel. The plant coefficients and the constant input delay are assumed known. The controller uses full-state feedback of the ODE and PIDE states together with the input history.

\begin{remark}
  As pointed out in \cite{krstic2008backstepping}, the open-loop system \eqref{xt}-\eqref{t-d} (with $U \left( t \right) =0$) may become unstable when the spatially varying coefficient $m(x)$ and the kernel function $f(x,y)$ of the integral term are sufficiently large.  
\end{remark}

\begin{remark}
Related PDE--ODE models arise in fermentation, stem cell dynamics, and flexible-structure control~\cite{schenk2022existence,batool2021well,chen2023rapid,mei2023exponential}, with either distributed interconnection or boundary coupling. These applications motivate the present study, although applying the proposed controller to their full models requires further modeling and analysis to verify that the stated assumptions hold.
\end{remark}

To compensate for the input delay in \eqref{t-d}, we introduce the following first-order hyperbolic (transport) PDE to model the delay:
\begin{align}
Dv_{t}(x, t) &= v_{x}(x, t),\label{vt1} \\
v(1, t) &= U(t),\label{vt2}
\end{align}
where $v:[0,1] \times[0,+\infty) \rightarrow \mathbb{R} $. In this case, the solution to \eqref{vt1}-\eqref{vt2} is $v(x, t) = U(t + Dx- D)$, and therefore we can further obtain 
\begin{align}
    U(t-D)=v(0,t).\label{ud}
\end{align}
Based on \eqref{vt1}-\eqref{ud}, the original system \eqref{xt}-\eqref{t-d} is transformed into the following intermediate system
\begin{subequations}
\begin{align}
\dot{X}(t) &=AX(t)+\int_{0}^{1} \mathcal{B}(x)u(x,t)dx, \label{disx}\\
u_{t}(x,t) &= \lambda u_{x}(x,t)+m(x)u(0,t)+\int_{0}^{x} f(x,y)u(y,t)dy, \\
u(1,t) &=v(0, t), \\
Dv_{t}(x, t) &= v_{x}(x, t), \\
v(1, t) &= U(t).\label{interv}
\end{align}
\end{subequations}

\begin{remark}
For every prescribed finite known constant delay $D>0$, the objective is to exponentially stabilize the augmented state $(X,u,v)$ using a causal dynamic autoregressive boundary feedback. The controller uses the current ODE and PIDE states together with the past control inputs represented by $v$.
\end{remark}

\begin{figure*}
    \centering
    \includegraphics[width=6in]{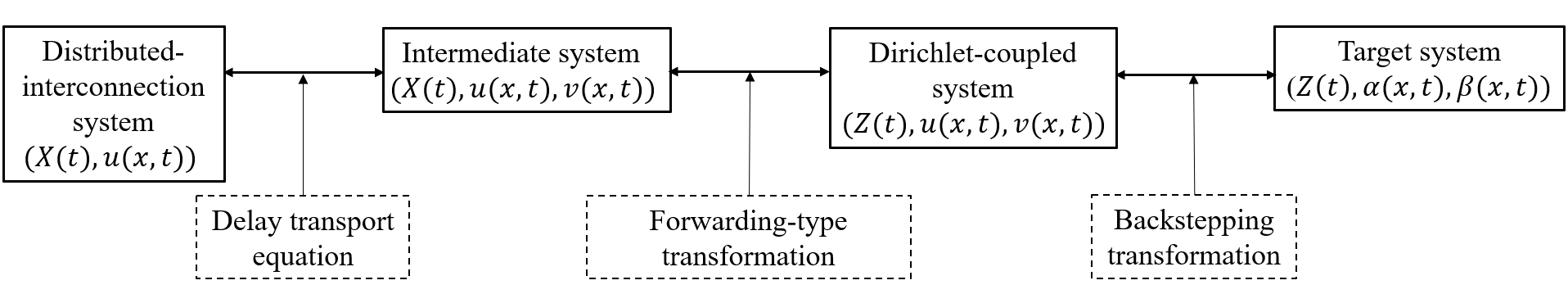}
    \caption{Process flowchart for PDE-ODE systems with distributed interconnection and an arbitrarily large delay}
    \label{liuchengtu}
\end{figure*}

However, the distributed interconnection arising from the cascade PIDE-ODE system in \eqref{disx} presents significant challenges for both controller design and stability analysis, because the ODE in \eqref{disx} depends on the PIDE state over the entire spatial domain. We first convert this distributed interconnection into boundary coupling. The main results are summarized in the next section, with the specific research steps outlined in the workflow depicted in Fig.\ref{liuchengtu}.

\section{Main Results}\label{3}
To reduce the distributed interconnection, we first introduce a forwarding-type integral transformation that converts the distributed PIDE--ODE coupling into an equivalent Dirichlet-coupled cascade. Based on this reformulation, a backstepping-based controller is then developed. Finally, the stability of the resulting closed-loop system is  analyzed.

\subsection{Distributed-to-Boundary Reduction}

To reduce the distributed interconnection to a boundary coupling, we introduce the forwarding-type transformation
\begin{align}
    Z(t)=X(t)+\int_{0}^{1} g(x)u(x,t)dx,\label{ODEtrans}
\end{align}
where $g(x)$ is the solution to the following boundary value problem
\begin{align}
\lambda g^{\prime}(x)&=\mathcal{B}(x)-A g(x)+\int_{x}^{1} g(y) f(y,x) dy \label{g'},\\
g(1)&=0. \label{gbound}
\end{align}
Differentiating \eqref{ODEtrans}, integrating by parts, and using \eqref{g'}--\eqref{gbound}, we obtain
\begin{align}
\dot Z(t)
={}&AZ(t)+\lambda g(1)u(1,t)\notag\\
&+\big[\int_0^1g(x)m(x)\,dx-\lambda g(0)\big]u(0,t)\notag\\
&+\int_0^1\big[\mathcal B(x)-Ag(x)-\lambda g'(x)\notag\\
&
+\int_x^1g(y)f(y,x)\,dy\big]u(x,t)\,dx\notag\\
={}&AZ(t)+Bu(0,t),
\label{zt}
\end{align}
where
\begin{align}
B=\int_0^1g(x)m(x)\,dx-\lambda g(0).
\label{B}
\end{align}

Based on the transformation \eqref{ODEtrans}, the dynamics can be written as the following PIDE-ODE Dirichlet-coupled system
\begin{subequations}
\begin{align}
\dot{Z}(t)&=AZ(t)+Bu(0,t),\\
u_{t}(x,t) &= \lambda u_{x}(x,t)+m(x)u(0,t) \notag \\&+\int_{0}^{x} f(x,y)u(y,t)dy, \\
Dv_{t}(x, t) &= v_{x}(x, t), \\
u(1,t) &=v(0, t), \\
v(1, t) &= U(t),
\end{align}
\end{subequations}
\begin{assu}\label{ass1}
Let \(B\) be defined by \eqref{B}. The pair \((A,B)\) is stabilizable.
\end{assu}

\begin{remark}
Assumption~\ref{ass1} can be checked using the original system coefficients. First, solve \eqref{g'}--\eqref{gbound} for $g(x)$, and the unique solvability of these equations is established in Theorem~\ref{thm:g}. Then compute $B$ from \eqref{B} and verify the stabilizability of the pair $(A,B)$ before selecting $K$.
\end{remark}

\begin{remark}
If $m(x)u(0,t)$ is replaced by $m(x)u(x_1,t)$  for some intermediate point $x_1\in(0,1)$, transformation~\eqref{ODEtrans} can still reduce the distributed interconnection to a boundary coupling, provided that the modified kernel problem is solvable. The kernel may have a jump at $x_1$, whose associated interface term cancels the contribution of $u(x_1,t)$ in the transformed ODE. However, the PIDE retains this term, so its stabilization and delay compensation require a modified backstepping design.
\end{remark}

The existence and uniqueness of the forwarding kernel $g$ are
established next.
\begin{theorem}\label{thm:g}
The integro-differential boundary value problem \eqref{g'}-\eqref{gbound}
admits a unique solution
$g\in C^{1}([0,1];\mathbb{R}^{n})$. 
Moreover, $\|g(x)\|\le \frac{\bar{\mathcal B}}{\lambda}e^{\frac{M}{\lambda}},
 x\in[0,1]$, where $M=\|A\|+\bar f, 
\bar f=\max_{(x,y)\in \mathcal T}|f(x,y)|, 
\bar{\mathcal B}=\max_{x\in[0,1]}\|\mathcal B(x)\|$.
\end{theorem}

\begin{proof}
Integrating \eqref{g'} from $x$ to $1$, using
\eqref{gbound}, and changing the order of integration give
the equivalent Volterra equation
\begin{align}
g(x)=F_0(x)+\int_x^1G(x,y)g(y)\,dy,
\label{eq:g-volterra}
\end{align}
where
\begin{align}
F_0(x)&=-\frac{1}{\lambda}
\int_x^1\mathcal B(s)\,ds,
\label{F}\\
G(x,y)&=\frac{1}{\lambda}
\left[A-I_n\int_x^y f(y,s)\,ds\right].
\label{G}
\end{align}

Following the successive-approximation argument used in the proof of Theorem~1 in~\cite{krstic2008backstepping}, set $\Delta g_0=F_0$ and define
\begin{align}
\Delta g_{n+1}(x)
=\int_x^1G(x,y)\Delta g_n(y)\,dy,
\qquad n\ge0.
\label{deltag}
\end{align}
Since $\|F_0(x)\|\le\bar{\mathcal B}/\lambda$
and $\|G(x,y)\|\le M/\lambda$, induction gives
\begin{align}
\|\Delta g_n(x)\|
\le \frac{\bar{\mathcal B}}{\lambda}
\frac{[M(1-x)/\lambda]^n}{n!}.
\label{induction}
\end{align}
Hence, the series $g=\sum_{n=0}^{\infty}\Delta g_n$
converges uniformly to a continuous solution of
\eqref{eq:g-volterra}, with
\begin{align}
\|g(x)\|
\le \sum_{n=0}^{\infty}\|\Delta g_n(x)\|
\le \frac{\bar{\mathcal B}}{\lambda}e^{M/\lambda}.
\end{align}
Since $F_0$ and $G$ have continuous derivatives with respect
to $x$, differentiation of \eqref{eq:g-volterra} yields
\eqref{g'}. Thus, $g\in C^1([0,1];\mathbb R^n)$.

To prove uniqueness, let $g_1$ and $g_2$ be two solutions of \eqref{g'}--\eqref{gbound}, and set $h=g_1-g_2$. Their difference satisfies
\[
h(x)=\int_x^1G(x,y)h(y)\,dy.
\]
Since $\|G(x,y)\|\le c$, where
$c=(\|A\|+\bar f)/\lambda$, it follows that
\[
\|h(x)\|\le c\int_x^1\|h(y)\|\,dy.
\]
Define $H(x)=\int_x^1\|h(y)\|\,dy$.
Then $H'(x)+cH(x)\ge0$ and $H(1)=0$.
Hence $e^{cx}H(x)$ is nondecreasing, and
\[
0\le e^{cx}H(x)\le e^cH(1)=0,
\qquad 0\le x\le1.
\]
Therefore, $H\equiv0$ and $h\equiv0$, proving uniqueness.

This completes the proof.
\end{proof}

Thus, the original system \eqref{xt}--\eqref{t-d} is transformed into the following equivalent PIDE--ODE cascade, where the ODE is driven by the PIDE boundary value $u(0,t)$:
\begin{subequations}
\begin{align}
\dot{Z}(t) &=A Z(t)+B u(0, \mathrm{t}), \label{origin1}\\
    u_{t} &= \lambda u_{x}+m(x)u(0,t)+\int_{0}^{x} f(x,y)u(y,t)dy, \\
Dv_{t}(x, t) &= v_{x}(x, t), \\
u(1,t) &=v(0, t), \\
v(1, t) &= U(t),\label{origin2}
\end{align}
\end{subequations}
Under Assumption \ref{ass1}, choose \(K\in\mathbb R^{1\times n}\) such that \(A+BK\) is Hurwitz.

\subsection{Backstepping-based Controller Design}
Following the backstepping approach \cite{krstic2008backstepping}, we introduce two transformations given by
\begin{align} 
\alpha(x, t) &=u(x, t)-\int_{0}^{x} K^{11}(x, y) u(y, t) d y-\gamma(x) Z(t), \label{transformation1} \\
\beta(x, t) &=v(x, t)-\int_{0}^{1} K^{21}(x, y) u(y, t) d y \notag \\ 
 &-\int_{0}^{x} K^{22}(x, y) v(y, t) d y-\delta(x) Z(t).\label{transformation2} 
\end{align} 
The kernel functions in \eqref{transformation1}-\eqref{transformation2} must satisfy the following PDEs and ODEs: 
\begin{subequations}
\begin{align}
\lambda K^{11}_{x}(x, y)
&=-\lambda K^{11}_{y}(x, y)+\int_{y}^{x} K^{11}(x, \xi) f(\xi, y)\,d \xi -f(x, y),
\label{k1xy} \\
\lambda K^{11}(x, 0)&=\int_{0}^{x} m(y) K^{11}(x, y)\,d y+\gamma(x)B-m(x),
\label{k10}\\
K^{21}_{x}(x, y)
&=-D\lambda K^{21}_{y}(x, y)+D\int_{y}^{1}K^{21}(x,\xi)f(\xi,y)\,d\xi,
\label{k2xy}\\
K^{21}(0,y)&=K^{11}(1,y),
\label{k2bd}\\
\lambda K^{21}(x, 0)&=\int_{0}^{1} m(y) K^{21}(x, y)\,d y+\delta(x)B,
\label{k2x0}\\
K^{22}_{x}(x, y)&=-K^{22}_{y}(x, y),
\label{k3xy}\\
K^{22}(x, 0)&=D\lambda K^{21}(x, 1),
\label{k3x0}\\
\lambda \gamma'(x)&=\gamma(x)A,
\label{kernel1}\\
\gamma(1)&=\delta(0),
\label{k1}\\
\delta'(x)&=D\delta(x)A,
\label{kernel2}\\
\gamma(0)&=K.
\label{k2}
\end{align}
\end{subequations}

Let $a=D\lambda$ and define
\begin{align}
\Sigma&=\{(x,y)\in[0,1]^2:y=ax\},\notag\\
\Omega_-&=\{(x,y)\in[0,1]^2:y<ax\},\notag\\
\Omega_+&=\{(x,y)\in[0,1]^2:y>ax\}.
\end{align}
The following theorem establishes the existence and uniqueness of the transformation kernels.

\begin{theorem}\label{thm:kernel}
At incompatible corners, the boundary conditions are interpreted through the corresponding one-sided traces. For every fixed $D>0$, the kernel system \eqref{k1xy}--\eqref{k2} admits a unique bounded solution. The functions $\gamma$, $\delta$, and $K^{11}$ are continuously differentiable, while $K^{21}$ and $K^{22}$ are piecewise continuously differentiable.

The kernel $K^{21}$ is given by
\[
K^{21}(x,y)=
\begin{cases}
K^{21,-}(x,y), & (x,y)\in\Omega_-,\\
K^{21,+}(x,y), & (x,y)\in\Omega_+,
\end{cases}
\]
where both branches are continuous on their respective domains. Their one-sided traces across $\Sigma$ satisfy
\begin{align}
K^{21,-}(x,ax)-K^{21,+}(x,ax)=\frac{m(1)}{\lambda},
0\le x\le\min\{1,a^{-1}\}.
\label{k2jump}
\end{align}

The kernel $K^{22}$ is determined by
\begin{equation}
K^{22}(x,y)=aK^{21}(x-y,1),
\qquad (x,y)\in\mathcal T.
\label{k22formula}
\end{equation}
If $a>1$, its one-sided traces across
$\Gamma_0=\{(x,y)\in\mathcal T:\ x-y=a^{-1}\}$
satisfy
\begin{equation}
K^{22,-}(x,y)-K^{22,+}(x,y)=Dm(1),
\qquad (x,y)\in\Gamma_0,
\label{k22jump}
\end{equation}
where the minus and plus traces correspond to $x-y>a^{-1}$ and $x-y<a^{-1}$, respectively. If $a\le1$, $K^{22}$ has no interior value jump.
\end{theorem}

\begin{proof}
See the appendix.
\end{proof}

Differentiate \eqref{transformation1} and \eqref{transformation2} with respect to space and time, and substitute the resulting expressions into the system \eqref{origin1}-\eqref{origin2}. The control law can be derived as
\begin{align} \label{controller}
v(1,t)=U(t)&=\int_{0}^{1} K^{21}(1, y) u(y, t) d y \notag \\
 &+\int_{0}^{1} K^{22}(1, y) v(y, t) d y+\delta(1) Z(t). 
\end{align}

Using the forwarding-type transformation \eqref{ODEtrans}, the controller
\eqref{controller} can be equivalently expressed in terms of the original
state variables as
\begin{align}
U(t)
&=\delta(1)X(t)
+\int_0^1 \bar K_u(y)u(y,t)\,dy\notag \\
&+\int_0^1K^{22}(1,y)v(y,t)\,dy,
\label{controller-original}
\end{align}
where
\begin{align}
\bar K_u(y)=K^{21}(1,y)+\delta(1)g(y).
\end{align}

Thus, the target system can be written as
\begin{subequations}
\begin{align} \label{target11}
\dot{Z}(t) &=(A+B K) Z(t)+B \alpha(0, t), \\
\alpha_{t}(x, t) &=\lambda \alpha_{x}(x, t), \\
D \beta_{t}(x, t) &=\beta_{x}(x, t),\label{tabeta} \\
\alpha(1, t) &=\beta(0, t), \\
\beta(1, t) &=0, \label{target22}
\end{align}
\end{subequations}
where $(A+BK)$ is Hurwitz.

\newtheorem{lemma}{Lemma}

\begin{lemma}\label{lem:jump-cancel}
Under the assumptions of Theorem~\ref{thm:kernel},
the interface terms associated with the jump of
$K^{21}$ across $\Sigma$ and, when $D\lambda>1$,
the jump of $K^{22}$ across $\Gamma_0$ cancel 
in the derivation of
\[
D\beta_t(x,t)-\beta_x(x,t)=0.
\]
Consequently, transformations
\eqref{transformation1}--\eqref{transformation2},
under the feedback law~\eqref{controller},
map system~\eqref{origin1}--\eqref{origin2}
into the target system~\eqref{target11}--\eqref{target22}.
\end{lemma}

\begin{proof}
    See the appendix.
\end{proof}

\subsection{Stability analysis}
This section establishes exponential stability of the original closed-loop system by proving bounded invertibility of the transformations and exponential stability of the target system.

We begin by presenting the inverse transformation that maps the target system states $(Z(t)$$,$ $\alpha (x,t)$$,$ $\beta(x,t))$ back to the original system states $(X(t)$$,$ $u(x,t)$$,$ $v(x,t))$. The inverse transform corresponding to \eqref{transformation1}-\eqref{transformation2} is given by 
\begin{align}
    u(x, t) & =\alpha(x, t)+\int_{0}^{x} H_{1}(x, y) \alpha(y, t) d y+\theta_1 (x) Z(t), \\
v(x, t) & =\beta(x, t)+\int_{0}^{1} H_{2}(x, y) \alpha(y, t) d y \notag\\
& +\int_{0}^{x} H_{3}(x, y) \beta(y, t) d y+\theta_2(x) Z(t).
\end{align}

Substitution into the system equations and integration by parts give the following inverse kernel equations
\begin{subequations}
\begin{align}
\lambda H_{1,x}(x,y)+\lambda H_{1,y}(x,y)
&=-f(x,y) \notag \\ 
&-\int_y^x f(x,\xi)H_1(\xi,y)\,d\xi,  \label{h1}\\
\frac{1}{D}H_{2,x}(x,y)+\lambda H_{2,y}(x,y)
&=0,  \label{b}\\
H_{3,x}(x,y)+H_{3,y}(x,y)
&=0, \label{c}\\
\lambda \theta_1'(x)-\theta_1(x)(A+BK)
&=-m(x)\theta_1(0)\notag \\
&-\int_0^x f(x,y)\theta_1(y)\,dy,  \label{d}\\
\frac{1}{D}\theta_2'(x)-\theta_2(x)(A+BK)
&=0,  \label{the2}\\
\theta_1(x)B-\lambda H_1(x,0)
&=m(x),  \label{f}\\
\lambda H_2(x,1)-\frac{1}{D}H_3(x,0)
&=0,  \label{g}\\
\lambda H_2(x,0)-\theta_2(x)B
&=0,  \label{h}\\
\theta_1(0)&=K,  \label{i}\\
H_1(1,y)&=H_2(0,y),  \label{j}\\
\theta_1(1)&=\theta_2(0).\label{the}
\end{align}
\end{subequations}

For each fixed $D>0$, Theorems~\ref{thm:g}
and~\ref{thm:kernel} ensure that the transformation
kernels are bounded.
Given $(Z,\alpha,\beta)$, equation~\eqref{transformation1}
is a Volterra integral equation of the second kind
that uniquely determines $u$.
With $u$ known, equation~\eqref{transformation2}
similarly determines $v$, and \eqref{ODEtrans}
then recovers $X$.
The bounded Volterra kernels $K^{11}$ and $K^{22}$
yield bounded inverse operators on $L^2(0,1)$,
while the integral involving $K^{21}$ defines
a bounded operator despite its discontinuities.
Consequently, the complete transformation is boundedly invertible on $\mathbb R^n\times L^2(0,1)\times L^2(0,1)$. 
For every initial state in $\mathcal H=\mathbb R^n\times L^2(0,1)\times L^2(0,1)$, the closed-loop system has a unique solution in $C([0,\infty);\mathcal H)$ that depends continuously on the initial state. The boundary signals are defined by the characteristic formulas.
The original and target state norms are therefore equivalent, allowing the exponential stability estimate for the target system to be transferred to the original closed-loop system.

Next, the stability of the target system \eqref{target11}-\eqref{target22} is established by the following theorem.

\begin{theorem}
Consider the closed-loop system \eqref{target11}-\eqref{target22}.
For any initial condition $(Z(0), \alpha(\cdot , 0), \beta(\cdot,0)) \in(  \mathbb{R}^{n} \times L^2(0,1) \times L^2(0,1))$,
the closed-loop system is exponentially stable in the sense that there exist positive constants $\mathcal M$ and $\kappa$, such that 
\begin{align}
\Omega(t)\le \mathcal M \Omega(0)e^{-\kappa t},  t\ge 0,
\end{align}
where
\begin{align}
\Omega(t)
=\|\alpha(\cdot,t)\|_{L^2}^2
+\|\beta(\cdot,t)\|_{L^2}^2
+\|Z(t)\|^2.
\end{align}
\end{theorem}

\begin{proof}
By the method of characteristics, $\beta(\cdot,t)=0$ for
$t\ge D$ and $\alpha(\cdot,t)=0$ for
$t\ge T_*=D+1/\lambda$. Thereafter, the target ODE reduces
to $\dot Z=(A+BK)Z$, where $A+BK$ is Hurwitz.
Together with a bound on the transient response over $[0,T_*]$,
this observation yields exponential stability of the target system.
We use the following Lyapunov functional to obtain an explicit
estimate for the complete target state for all $t\ge0$.

Let $A_c=A+BK$. Since $A_c$ is Hurwitz, there exists a symmetric positive definite matrix $P$ such that
\begin{align}
A_c^\top P+PA_c=-I_n.
\label{eq:lyap-matrix}
\end{align}
Choose $\phi>4\|PB\|^2/\lambda$ and define
\begin{align}
V(t)={}&Z^\top(t)PZ(t)
+\frac{\phi}{2}\int_0^1(1+x)\alpha^2(x,t)\,dx \notag\\
&
+\phi\lambda D\int_0^1(1+x)\beta^2(x,t)\,dx.
\label{eq:lyap-functional}
\end{align}

For regular initial conditions satisfying the boundary conditions, differentiation along \eqref{target11}--\eqref{target22} and integration by parts yields
\begin{align}
\dot V={}&-\|Z\|^2+2Z^\top PB\alpha(0,t)
-\frac{\phi\lambda}{2}|\alpha(0,t)|^2
-\frac{\phi\lambda}{2}\|\alpha(\cdot,t)\|_{L^2}^2\notag\\
&
-\phi\lambda\|\beta(\cdot,t)\|_{L^2}^2.
\end{align}
Here, the boundary terms $\phi\lambda\alpha^2(1,t)$ and $-\phi\lambda\beta^2(0,t)$ cancel because $\alpha(1,t)=\beta(0,t)$, while $\beta(1,t)=0$.

By Young's inequality,
\begin{align}
2Z^\top PB\alpha(0,t)
\le \frac12\|Z\|^2+2\|PB\|^2|\alpha(0,t)|^2.
\end{align}
The choice of $\phi$ therefore gives
\begin{align}
\dot V
&\le -\frac12\|Z\|^2
-\frac{\phi\lambda}{2}\|\alpha(\cdot,t)\|_{L^2}^2
-\phi\lambda\|\beta(\cdot,t)\|_{L^2}^2
\le -\kappa V,
\label{eq:lyap-decay}
\end{align}
where
\begin{align}
\kappa=
\min\left\{
\frac{1}{2\lambda_{\max}(P)},
\frac{\lambda}{2},
\frac{1}{2D}
\right\}>0.
\end{align}
The last inequality follows from $Z^\top PZ\le\lambda_{\max}(P)\|Z\|^2$ and $1\le1+x\le2$.

For each fixed $D>0$, positivity of $P$ and the weights in \eqref{eq:lyap-functional} gives constants $c_1,c_2>0$, independent of the initial conditions, such that
\begin{align}
c_1\Omega(t)\le V(t)\le c_2\Omega(t).
\end{align}
Consequently, integrating \eqref{eq:lyap-decay} gives
\begin{align}
\Omega(t)
\le \frac{c_2}{c_1}e^{-\kappa t}\Omega(0),
\qquad t\ge0.
\label{eq:lyap-estimate}
\end{align}

For square-integrable initial functions, approximate the initial
state by regular data satisfying the target boundary conditions.
Continuous dependence of the target solution on the initial state
allows estimate~\eqref{eq:lyap-estimate} to pass to the limit on
each finite time interval. Hence, the same estimate holds for every
initial state in
$\mathbb R^n\times L^2(0,1)\times L^2(0,1)$,
with $\mathcal M=c_2/c_1$. This completes the proof.
\end{proof}

Finally, by the equivalence between the original closed-loop system \eqref{xt}-\eqref{t-d} under controller \eqref{controller} and the target system \eqref{target11}-\eqref{target22}, the following stability result holds.

\begin{theorem}
Consider the system \eqref{xt}-\eqref{t-d} under controller \eqref{controller} or equivalently its intermediate representation \eqref{disx}-\eqref{interv}. For any initial condition $(X(0), u(\cdot , 0), v(\cdot,0)) \in(  \mathbb{R}^{n} \times L^2(0,1) \times L^2(0,1))$, the closed-loop system is exponentially stable in the sense that there exist positive constants $\tilde{c}$ and $\sigma $ such that  
\begin{align}
\Psi (t)\leq \tilde{c} e^{-\sigma  t}\Psi (0),
\end{align}
where 
\begin{align}
\Psi (t)=\|u(\cdot,t) \|_{L^{2}}^{2}+\|v(\cdot,t) \|_{L^{2}}^{2}+\| X(t) \|^{2}.
\end{align}
\end{theorem}

\begin{proof}
The transformation identities and the feedback boundary condition
map the original closed-loop system into the target system.
For each fixed $D>0$, bounded invertibility of the complete
transformation gives constants $c_-,c_+>0$, independent of time
and the initial conditions, such that
\begin{align}
c_-\Psi(t)\le\Omega(t)\le c_+\Psi(t).
\label{eq:state-norm-equivalence}
\end{align}
Combining this inequality with the estimate of Theorem~3 gives
\begin{align}
\Psi(t)
\le \frac{1}{c_-}\Omega(t) \le \frac{\mathcal M}{c_-}e^{-\kappa t}\Omega(0) \le \frac{\mathcal M c_+}{c_-}e^{-\kappa t}\Psi(0).
\label{eq:original-state-decay}
\end{align}
The conclusion follows with
$\tilde c=\mathcal M c_+/c_-$ and $\sigma=\kappa$.
\end{proof}

\begin{remark}
For a fixed plant and a stabilizing gain $K$ selected independently of $D$, the quantities $g$, $B$, $K^{11}$, and $\gamma$ are independent of $D$, whereas $\delta$, $K^{21}$, and $K^{22}$ depend on $D$ and must be updated whenever the prescribed delay changes. The controller ensures closed-loop exponential stability for every prescribed finite $D>0$, but the kernel and feedback bounds need not be uniform in $D$.
\end{remark}

\section{Simulation}\label{4}

\begin{figure}
    \centering
    \begin{subfigure}[b]{0.49\linewidth}
        \centering
        \includegraphics[width=\linewidth]{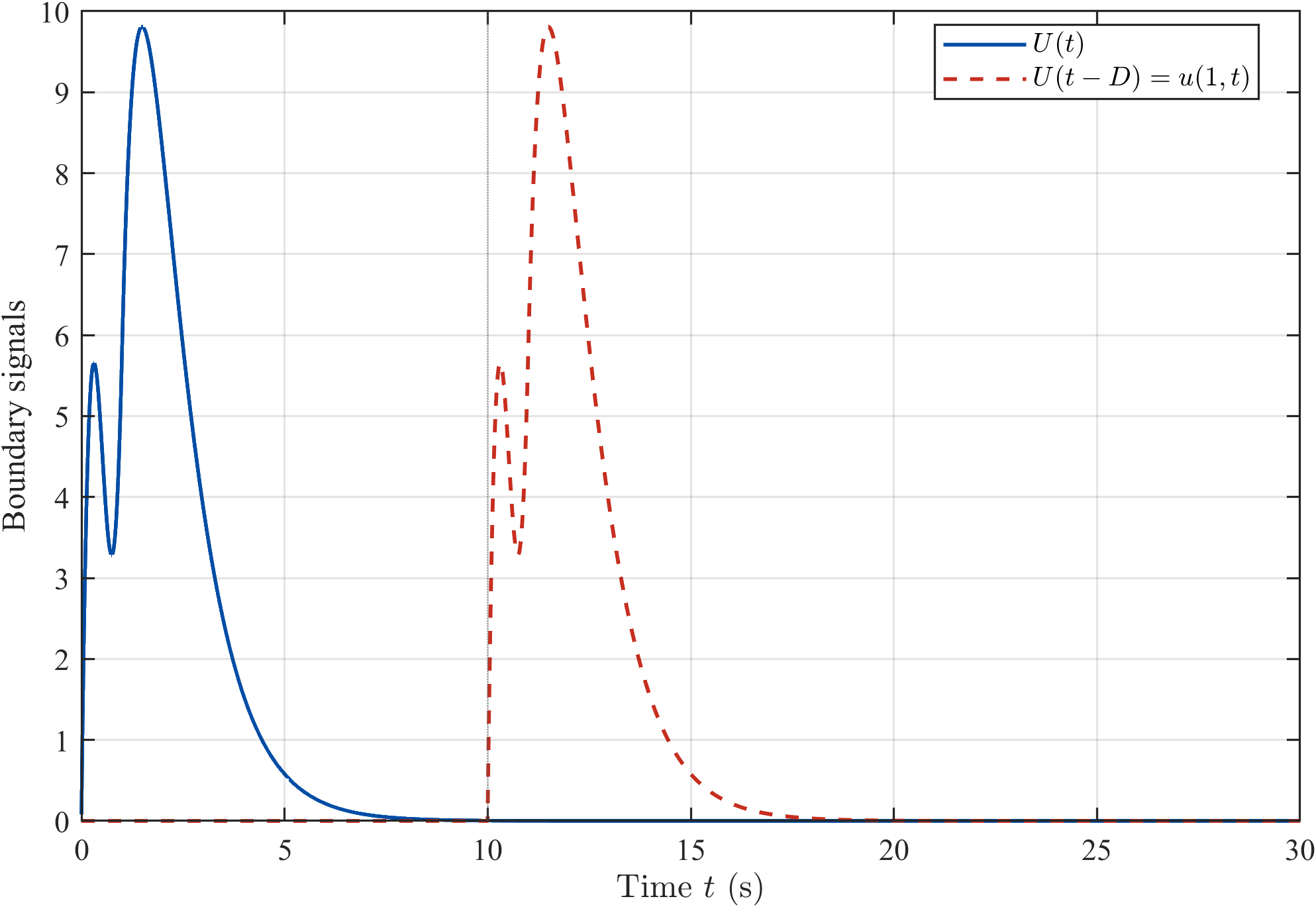}
        \caption{Control input $U(t)$ and delayed input $U(t-D)$.}
        \label{u}
    \end{subfigure}
    \hfill
    \begin{subfigure}[b]{0.49\linewidth}
        \centering
        \includegraphics[width=\linewidth]{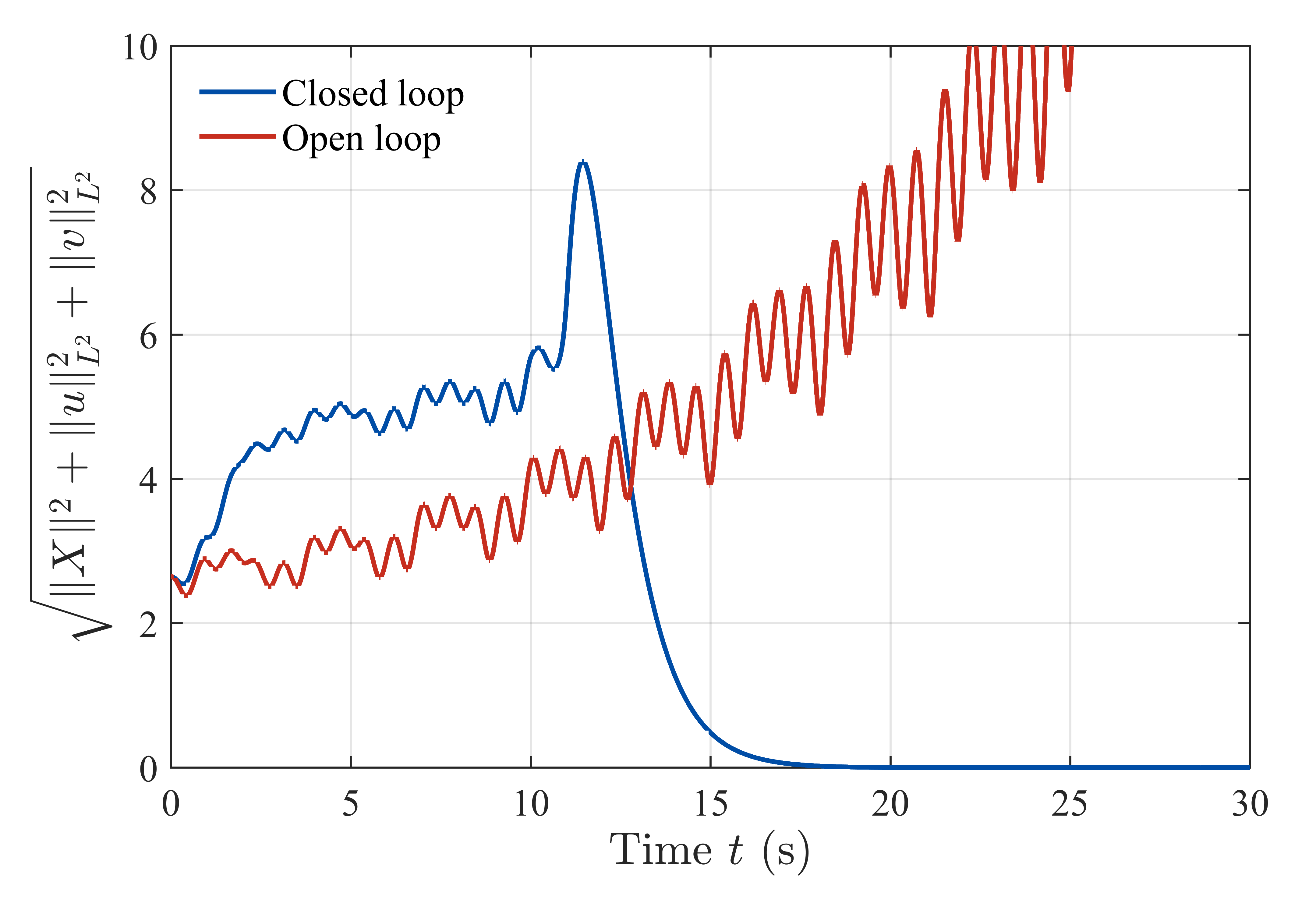}
        \caption{Open-loop and closed-loop state norms}
        \label{Z}
    \end{subfigure}
    \caption{Control inputs and state responses}
    \label{fig:sim}
\end{figure}

\begin{figure}
    \centering
    \begin{subfigure}[b]{0.49\linewidth}
        \centering
        \includegraphics[width=\linewidth]{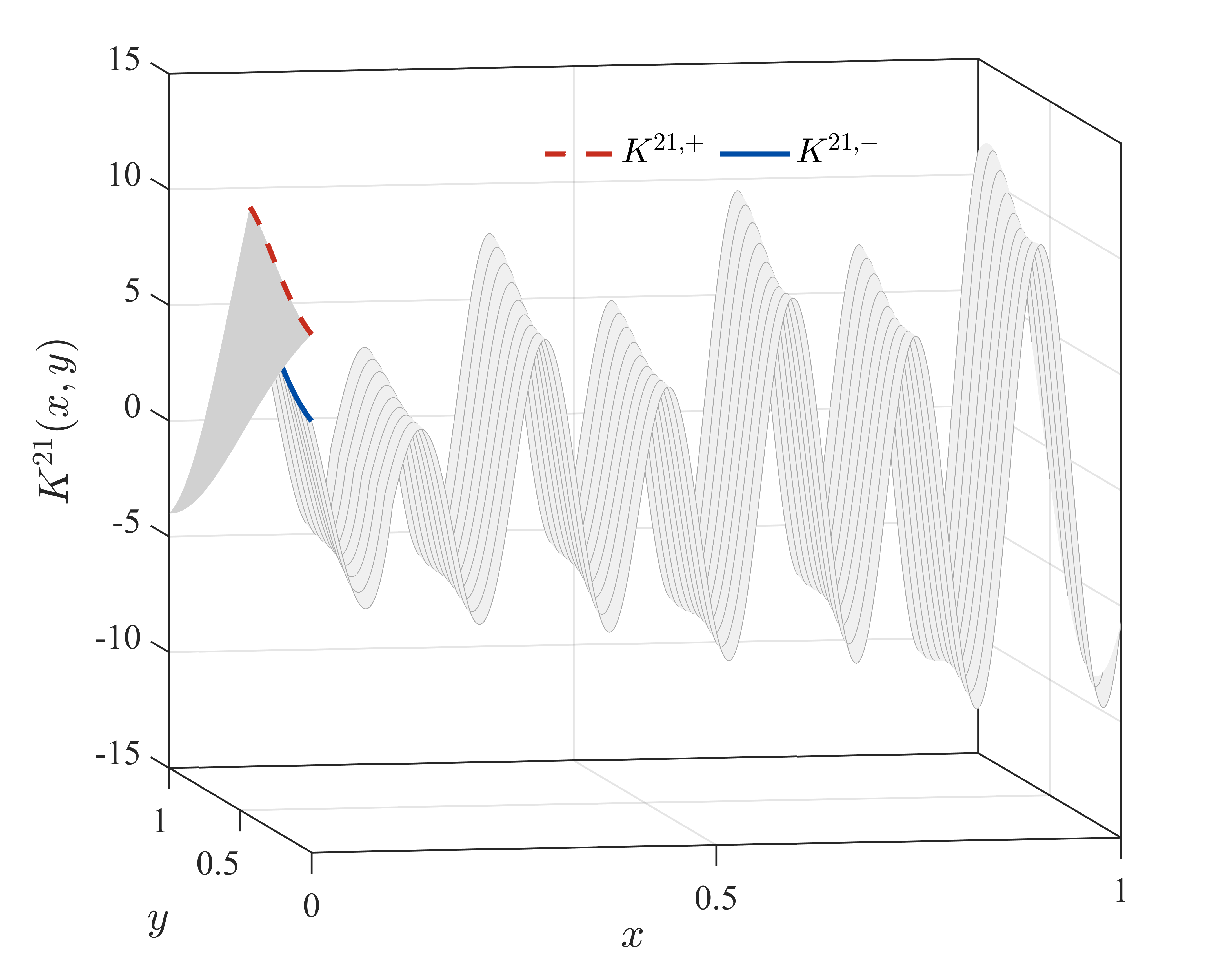}
        \caption{Discontinuous kernel $K^{21}$}
        \label{alpha1}
    \end{subfigure}
    \hfill
    \begin{subfigure}[b]{0.49\linewidth}
        \centering
        \includegraphics[width=\linewidth]{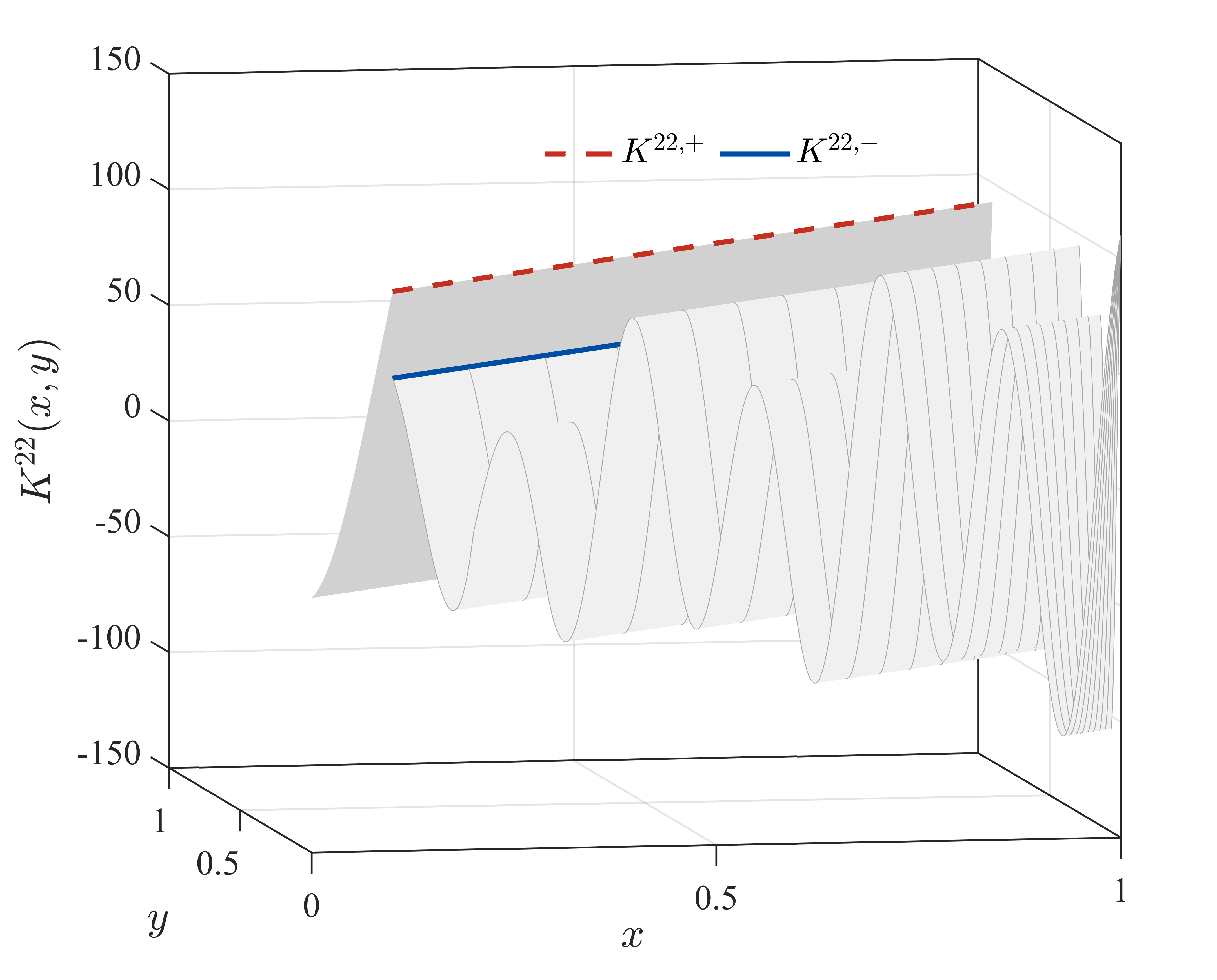}
        \caption{Discontinuous kernel $K^{22}$}
        \label{K21s1}
    \end{subfigure}
    \caption{Surface plots of the backstepping kernels}
    \label{fig:uxt_k21s1}
\end{figure}

\begin{figure}
    \centering
    \begin{subfigure}[b]{0.47\linewidth}
        \centering
        \includegraphics[width=\linewidth]{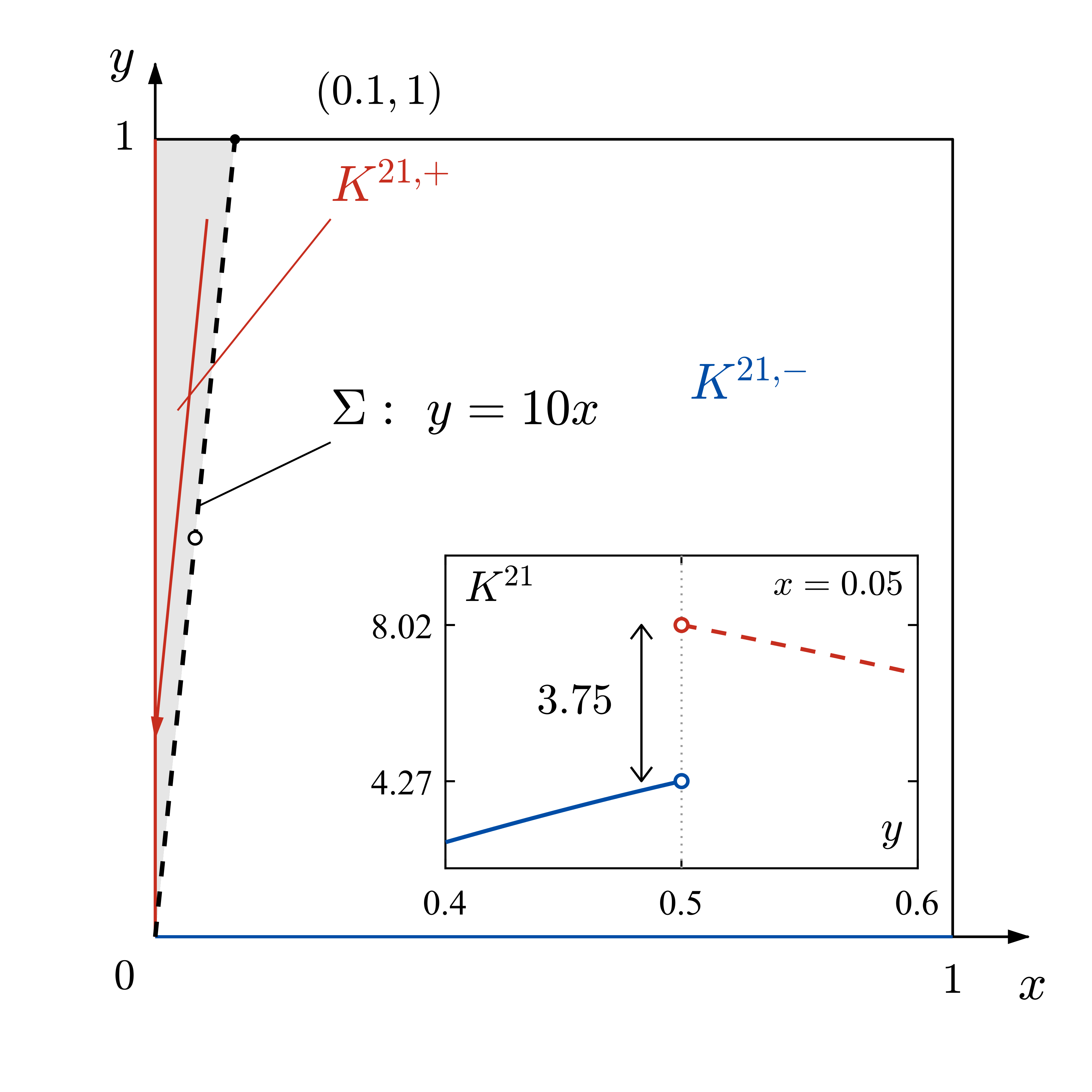}
        \caption{Domain partition of $K^{21}$}
        \label{K21}
    \end{subfigure}
    \hfill
    \begin{subfigure}[b]{0.47\linewidth}
        \centering
        \includegraphics[width=\linewidth]{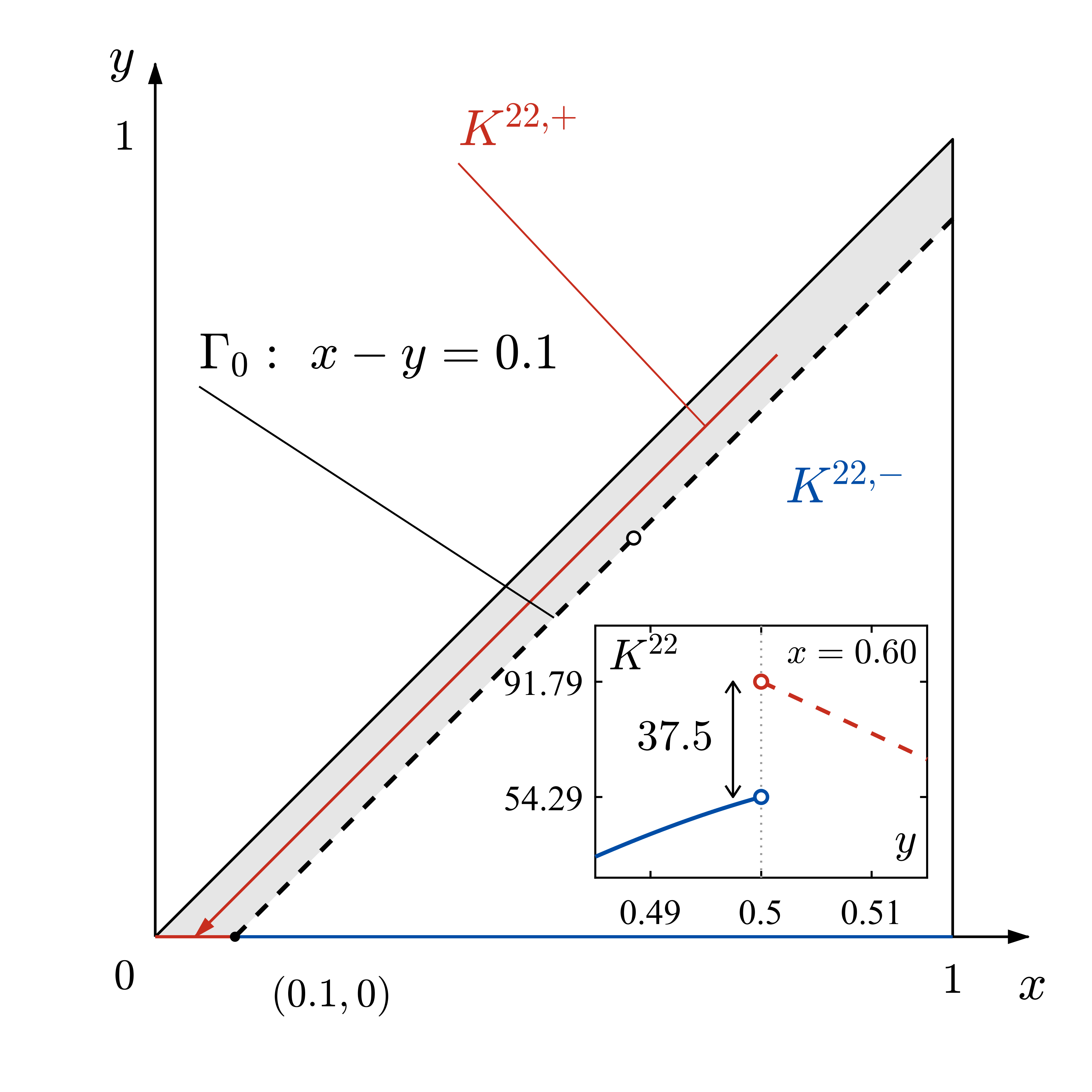}
        \caption{Domain partition of $K^{22}$}
        \label{K22}
    \end{subfigure}
    \caption{Characteristic partitions and jump interfaces of $K^{21}$ and $K^{22}$}
    \label{fig:kernels}
\end{figure}

A simulation example is presented to demonstrate the effectiveness of the proposed control design. Consider the following first-order hyperbolic PIDE-ODE system 
\begin{subequations}
\begin{align}
&\dot{X}(t)=\begin{bmatrix}
  0 & -2 \\
  2 & 0
\end{bmatrix} X(t)+\int_{0}^{1} \begin{bmatrix}
 -x\\x
\end{bmatrix}u(x,t)dx,\\
    &u_{t} = u_{x}-3.75xu(0,t)+\int_{0}^{x} (x+y)u(y,t)dy, \\
    &u(1,t) =U(t-10).
\end{align}\label{simu}
\end{subequations}
The initial conditions for system  \eqref{simu} are given by
\begin{align}
    &X(0)=\begin{bmatrix}
 1\\
2
\end{bmatrix}, u(x,0)=-2\sin(\pi x),x\in [0,1].
\end{align}
In addition, the input prehistory is chosen as $U(t)=0, t\in[-10,0)$, which is equivalently represented by the delay-state initial condition $v(x,0)=0, x\in[0,1]$.
The vector $B$ is computed from \eqref{B}, and the gain $K$ is selected by pole placement with desired poles at $-2$ and $-1$. Their rounded values are
\begin{align} \label{a+bk}
    B=\begin{bmatrix}
 -0.972\\
-0.143
\end{bmatrix},
K=\begin{bmatrix}3.17 & -0.56\end{bmatrix}.
\end{align}
Using the values reported in \eqref{a+bk}, the closed-loop matrix
$A+BK$ has eigenvalues approximately $-1.9976$ and $-1.0036$, and is therefore Hurwitz.

The controller~\eqref{controller} is then implemented.
Fig.~\ref{u} shows the control input $U(t)$ and the delayed
boundary input $U(t-D)$.
Fig.~\ref{Z} compares the open-loop and closed-loop responses
under the same initial conditions, using the total state norm.
The delayed control reaches the PIDE boundary at $t=D=10$.
The closed-loop norm subsequently decays to zero, whereas
the open-loop response exhibits growing oscillations.

Figs.~\ref{alpha1} and~\ref{K21s1} show the  surfaces of $K^{21}$ and $K^{22}$, respectively.
Figs.~\ref{K21} and~\ref{K22} show the corresponding
characteristic interfaces and numerical sections across
the jumps.
For $D=10$ and $\lambda=1$, one has $a=D\lambda=10$.
The jump interface of $K^{21}$ is therefore
$\Sigma:y=10x$.
Since $m(x)=-3.75x$, the signed jump defined
in~\eqref{k2jump} is $m(1)/\lambda=-3.75$,
with magnitude $3.75$.
The relation $K^{22}(x,y)=aK^{21}(x-y,1)$ maps the jump
of $K^{21}$ at $(0.1,1)$ to
$\Gamma_0:x-y=0.1$.
The signed jump of $K^{22}$ is $Dm(1)=-37.5$,
with magnitude $37.5$.

\section{Conclusion}\label{5}

This paper has developed a boundary feedback design for hyperbolic PIDE--ODE cascades with distributed interconnections and an arbitrarily large known constant input delay. 
A forwarding-type integral transformation reduces the distributed interconnection to an equivalent Dirichlet-coupled cascade, allowing subsequent stabilization and delay compensation without requiring the PIDE to be open-loop stable.
The kernel analysis identifies the characteristic propagation of incompatible boundary data and determines the resulting jump locations and amplitudes.
We prove that the interface terms cancel when deriving the target equations. Together with bounded invertibility of the transformation, this establishes closed-loop exponential stability despite the kernel discontinuities, without additional parameter restrictions to enforce kernel continuity.
These results provide a systematic basis for using discontinuous kernels in the considered delay-compensation design. The numerical example illustrates the effectiveness of the proposed controller.
Future research will consider uncertain delays and output-feedback extensions.

\section{Appendix} \label{app}
\subsection{Proof of Theorem \ref{thm:kernel}}
In this appendix, we provide the proof of Theorem \ref{thm:kernel}.

Equations \eqref{kernel1}--\eqref{k2} give
$\gamma(x)=Ke^{Ax/\lambda},
\delta(x)=\gamma(1)e^{DAx}.$
Thus, $\gamma$ and $\delta$ are uniquely determined.

First, we analyze the kernel $K^{11}(x,y)$.
Set $s=x-y$, $r=y$, and $q(s,r)=K^{11}(s+r,r)$.
Integration of \eqref{k1xy} along characteristics,
together with \eqref{k10}, yields
\begin{align}
q(s,r)
&=\frac{\gamma(s)B-m(s)}{\lambda}
-\frac{1}{\lambda}\int_0^r f(s+\rho,\rho)\,d\rho
\notag\\
&\quad+\frac{1}{\lambda}\int_0^s
m(\rho)q(s-\rho,\rho)\,d\rho
\notag\\
&\quad+\frac{1}{\lambda}\int_0^r\int_0^s
q(s-\eta,\rho+\eta)f(\rho+\eta,\rho)
\,d\eta\,d\rho.
\label{qfix}
\end{align}
The standard successive-approximation argument therefore gives a unique bounded solution. Since the coefficients are continuously differentiable, differentiation of \eqref{qfix} gives a Volterra equation with the same integral operator for $q_s$, while \eqref{k1xy} determines $q_r$. Hence $K^{11}(x,y)\in C^1(\mathcal T)$.

Second, we analyze the kernel $K^{21}(x,y)$.
Write $w=K^{21}$, $k(y)=K^{11}(1,y)$,
$d(x)=\delta(x)B/\lambda$, and $b(x)=w(x,0)$,
where $b(0)$ denotes the bottom-boundary trace at the corner.
The boundary condition \eqref{k2x0} becomes
\begin{equation}
b(x)=d(x)+\frac{1}{\lambda}
\int_0^1m(\eta)w(x,\eta)\,d\eta.
\label{wbottom}
\end{equation}
The characteristics of \eqref{k2xy} satisfy
$y-ax=\mathrm{constant}$. Using the bottom and left
boundary data on $\Omega_-$ and $\Omega_+$, respectively,
gives
\begin{align}
w_-(x,y)
&=b\!\left(x-\frac{y}{a}\right)
+D\int_{x-y/a}^{x}\int_{y+a(\sigma-x)}^1
w(\sigma,\zeta)
\notag\\
&\times f\bigl(\zeta,y+a(\sigma-x)\bigr)
\,d\zeta\,d\sigma,
\label{wminus}\\
w_+(x,y)
&=k(y-ax)
+D\int_0^x\int_{y-ax+a\sigma}^1
w(\sigma,\zeta)
\notag\\
&\times f\bigl(\zeta,y-ax+a\sigma\bigr)
\,d\zeta\,d\sigma.
\label{wplus}
\end{align}

Starting from $w^{(0)}=b^{(0)}=0$, compute $w^{(n+1)}$ from \eqref{wminus}--\eqref{wplus} using $w^{(n)}$ and $b^{(n)}$, and then update $b^{(n+1)}$ through \eqref{wbottom} using $w^{(n+1)}$. Let $W_n(x)$ and $F_n(x)$ denote the suprema of the absolute successive differences in $w$ and $b$, respectively, over first coordinates in $[0,x]$. The characteristic formulas give
\begin{align}
W_{n+1}(x)
&\le F_n(x)+C\int_0^x W_n(s)\,ds,\\
F_{n+1}(x)
&\le C\int_0^x\bigl(F_n(s)+W_n(s)\bigr)\,ds.
\end{align}
The second inequality follows by substituting \eqref{wminus}--\eqref{wplus} into \eqref{wbottom} and using the change of variables $s=x-\eta/a$ in the term containing $b^{(n)}$.

Set $E_n=W_n+F_n$. The above inequalities give $E_{n+1}(x)\le C_0E_n(x)$ and then we can get
\begin{align}
E_{n+2}(x)\le C\int_0^x E_n(s)\,ds,
\qquad n\ge1,
\end{align}
where the constants are independent of $n$. Consequently, for some $M>0$, we have
\begin{align}
E_{2j+1}(x)+E_{2j+2}(x)
\le M\frac{(Cx)^j}{j!},
\qquad j\ge0.
\end{align}
Hence, the iterates converge uniformly on each branch to a bounded solution of \eqref{wbottom}--\eqref{wplus}. Applying the same estimates to the difference of two solutions gives
\begin{align}
E(x)\le C\int_0^x E(s)\,ds.
\end{align}
Gronwall's inequality yields $E\equiv0$, proving uniqueness.

On $y=ax$, the integral terms in
\eqref{wminus} and \eqref{wplus} coincide. Therefore,
\begin{align}
w_-(x,ax)-w_+(x,ax)=b(0)-k(0).
\end{align}
Using \eqref{k10} at $x=1$ and the bottom-boundary
trace of \eqref{k2x0} at $x=0$, together with
\eqref{k2bd} and \eqref{k1}, yields
\begin{align}
\lambda\bigl(b(0)-k(0)\bigr)
=[\delta(0)-\gamma(1)]B+m(1)=m(1).
\end{align}
This proves \eqref{k2jump}. Differentiating the
characteristic integral equations after splitting
the spatial integrals at the interface gives
Volterra equations for the derivatives, including
the terms determined by \eqref{k2jump}.
The same successive-approximation argument establishes
piecewise continuous differentiability and one-sided
traces. 

Third, we analyze the kernel $K^{22}(x,y)$.
Integration of \eqref{k3xy} along $x-y=\mathrm{constant}$,
using \eqref{k3x0}, gives
\begin{align}
K^{22}(x,y)=aK^{21}(x-y,1),
\end{align}
which proves \eqref{k22formula} and boundedness of
$K^{22}(x,y)$. If $a>1$, the interface reaches $y=1$
at $x=a^{-1}$, so \eqref{k2jump} yields
\begin{align}
K^{22,-}(x,y)-K^{22,+}(x,y)
=a\frac{m(1)}{\lambda}=Dm(1),
\end{align}
on $\Gamma_0$, proving \eqref{k22jump}.
If $a<1$, the line $x-y=a^{-1}$ does not intersect $\mathcal T$; if $a=1$, it meets $\mathcal T$ only at $(1,0)$. Thus, $K^{22}(x,y)$ has no interior value jump when $a\le1$. Its piecewise continuous differentiability follows from \eqref{k22formula}. This completes the proof.

\subsection{Proof of Lemma \ref{lem:jump-cancel}} 
In this appendix, we provide the proof of Lemma \ref{lem:jump-cancel}.

By Theorem~\ref{thm:kernel}, the jump of $K^{21}(x,y)$ satisfies \eqref{k2jump}, and the jump of $K^{22}(x,y)$ satisfies \eqref{k22jump} when $a>1$.
We first consider regular compatible states and define
\begin{align}
I(x,t)&=\int_0^1K^{21}(x,y)u(y,t)\,dy,\\
J(x,t)&=\int_0^xK^{22}(x,y)v(y,t)\,dy.
\end{align}
For $0<ax<1$, split $I$ at $y=ax$:
\begin{align}
I={}&\int_0^{ax}K^{21,-}(x,y)u(y,t)\,dy +\int_{ax}^1K^{21,+}(x,y)u(y,t)\,dy.
\end{align}
Leibniz's rule and the trace relation~\eqref{k2jump} give
\begin{align}
I_x=\int_0^1K^{21}_x(x,y)u(y,t)\,dy
     +Dm(1)u(ax,t).
\end{align}
Substituting the PIDE equation, integrating its transport
term by parts on each branch, and exchanging the order
of integration in the nonlocal term yield
\begin{align}
DI_t={}&
aK^{21}(x,1)u(1,t)-aK^{21}(x,0)u(0,t)\notag\\
&-a\int_0^1K^{21}_y(x,y)u(y,t)\,dy
+Dm(1)u(ax,t)\notag\\
&+D u(0,t)\int_0^1K^{21}(x,y)m(y)\,dy\notag\\
&+D\int_0^1
 \left[\int_y^1K^{21}(x,\xi)f(\xi,y)\,d\xi\right]
 u(y,t)\,dy.
\end{align}
Thus, the interface terms cancel in $DI_t-I_x$.
Using~\eqref{k2xy} and~\eqref{k2x0}, we obtain
\begin{align}
DI_t-I_x
=aK^{21}(x,1)u(1,t)-D\delta(x)Bu(0,t).
\end{align}
For $ax>1$, there is no interior interface in $I$ and
the same identity holds without the two interface terms.

When $a>1$ and $x>a^{-1}$, set $b=x-a^{-1}$ and write
\begin{align}
J={}&\int_0^bK^{22,-}(x,y)v(y,t)\,dy+\int_b^xK^{22,+}(x,y)v(y,t)\,dy.
\end{align}
Since $b'(x)=1$ and $Dv_t=v_x$, using  Leibniz's rule,
branch-wise integration by parts, and~\eqref{k22jump} give
\begin{align}
J_x={}&K^{22}(x,x)v(x,t)+\int_0^xK^{22}_x(x,y)v(y,t)\,dy
+Dm(1)v(b,t),\\
DJ_t={}&K^{22}(x,x)v(x,t)-K^{22}(x,0)v(0,t)\notag\\
&-\int_0^xK^{22}_y(x,y)v(y,t)\,dy
+Dm(1)v(b,t).
\end{align}
The interface terms also cancel in $DJ_t-J_x$.
Equation~\eqref{k3xy} therefore implies
\begin{align}
DJ_t-J_x=-K^{22}(x,0)v(0,t).
\end{align}
For $a\le1$, or for $a>1$ and $x<a^{-1}$, the same
identity holds without interface terms.

Using $\beta=v-I-J-\delta Z$,
$\dot Z=AZ+Bu(0,t)$, and $u(1,t)=v(0,t)$, we obtain
\begin{align}
D\beta_t-\beta_x
={}&[K^{22}(x,0)-aK^{21}(x,1)]v(0,t)\notag\\
&+[\delta'(x)-D\delta(x)A]Z(t)=0,
\end{align}
where the last equality follows from~\eqref{k3x0}
and~\eqref{kernel2}.

The three cases are considered separately:
\begin{itemize}
\item
If $a<1$, $\Sigma$ ends at $(1,a)$. The integral $I$ is split at $y=ax$, while $J$ has no interior jump interface.
\item
If $a=1$, $\Sigma$ ends at $(1,1)$, and $\Gamma_0$ meets the kernel domain only at $(1,0)$. Thus, $J$ has no interior jump interface. The kernel boundary conditions at these corners are understood through the corresponding one-sided limits.
\item
If $a>1$, $\Sigma$ ends at $(a^{-1},1)$ and $\Gamma_0$ starts at $(a^{-1},0)$. The integral $I$ is split for $x<a^{-1}$, and $J$ is split for $x>a^{-1}$.
\end{itemize}

Since the kernels are bounded, the integrals over intervals of vanishing length tend to zero at the transition point.
The remaining contributions are continuous, so no additional Dirac terms arise.

Finally, \eqref{k2bd}, \eqref{k1}, $v(0,t)=u(1,t)$, and the feedback law~\eqref{controller} give the target boundary conditions, while differentiating~\eqref{transformation1} yields the remaining target equations.
This completes the proof.

\bibliographystyle{IEEEtran}
\bibliography{IEEEabrv,references}

\end{document}